\documentclass[twocolumn]{aastex701}
\usepackage[utf8]{inputenc}
\usepackage{caption}
\usepackage{subcaption}
\usepackage{colortbl}
\usepackage{verbatim}
\usepackage{amsmath}

\shorttitle{Nascent EBs}
\shortauthors{Frank and Moe}

\newcommand{\cs}{$\chi^2$\,}
\newcommand{\ks}{\,km\,s$^{-1}$\,}
\newcommand{\Lsun}{L$_\odot$}
\newcommand{\Rsun}{R$_\odot$}
\newcommand{\Msun}{M$_\odot$}
\newcommand{\vsini}{$v$\,sin\,$i$ }

\newcommand{\logg}{log\,$g$ }

\begin{document}

\title{One-Million-Year-Old Nascent Eclipsing Binaries in the LMC}

\author[0009-0002-2866-9788]{Megan G. Frank}
\affil{Department of Physics and Astronomy, University of Wyoming, 1000~E.~University~Ave., Dept.~3905, Laramie, WY 82071, USA}
\email{mfrank13@uwyo.edu}

\author[0000-0002-0870-6388]{Maxwell Moe}
\affil{Department of Physics and Astronomy, University of Wyoming, 1000~E.~University~Ave., Dept.~3905, Laramie, WY 82071, USA}
\email{mmoe2@uwyo.edu}


\begin{abstract}

The light curves of nascent eclipsing binaries (EBs) feature sinusoidal reflection effects as a hot B-type main-sequence (MS) primary irradiates the near side of a cool, low-mass, pre-MS companion. We obtained multi-epoch echelle spectra with the MIKE spectrograph on the 6.5m Magellan-Clay Telescope of three nascent EBs embedded in H\,II regions within the LMC. By fitting atmospheric models to the spectra, the primaries are found to be mid-B dwarfs ($T_{\rm eff,1}$ = 18,600\,-\,22,800\,K; $M_1$~=~6.8\,-\,8.5\,\Msun).  The dynamical masses of the secondaries are only $M_2$ = 0.38\,-\,0.95\,\Msun, yielding extremely small mass ratios $q$ = $M_2$/$M_1$ = 0.05\,-\,0.11. Unlike solar-type primaries, which exhibit a dearth of close companions with $q$ $<$ 0.1 known as the brown dwarf desert, massive stars can readily form with very close companions below $q$ $<$ 0.1. By fitting PHOEBE models to the OGLE-III light curves, we measure the secondary radii $R_2$ = 2.0\,-\,2.8\,\Rsun\ and temperatures $T_{\rm eff,2}$ $\approx$ 4,000\,K. The pre-MS companions are located high on the Hayashi track where evolutionary models indicate ages of only $\approx$\,1 Myr. Their extreme youth suggests that nascent EBs formed via inward disk migration during the previous embedded protostellar phase.

\end{abstract}

\section{Introduction}\label{sec:Intro}

\begin{figure}[htp]
    \centering
    \includegraphics[width=\linewidth]{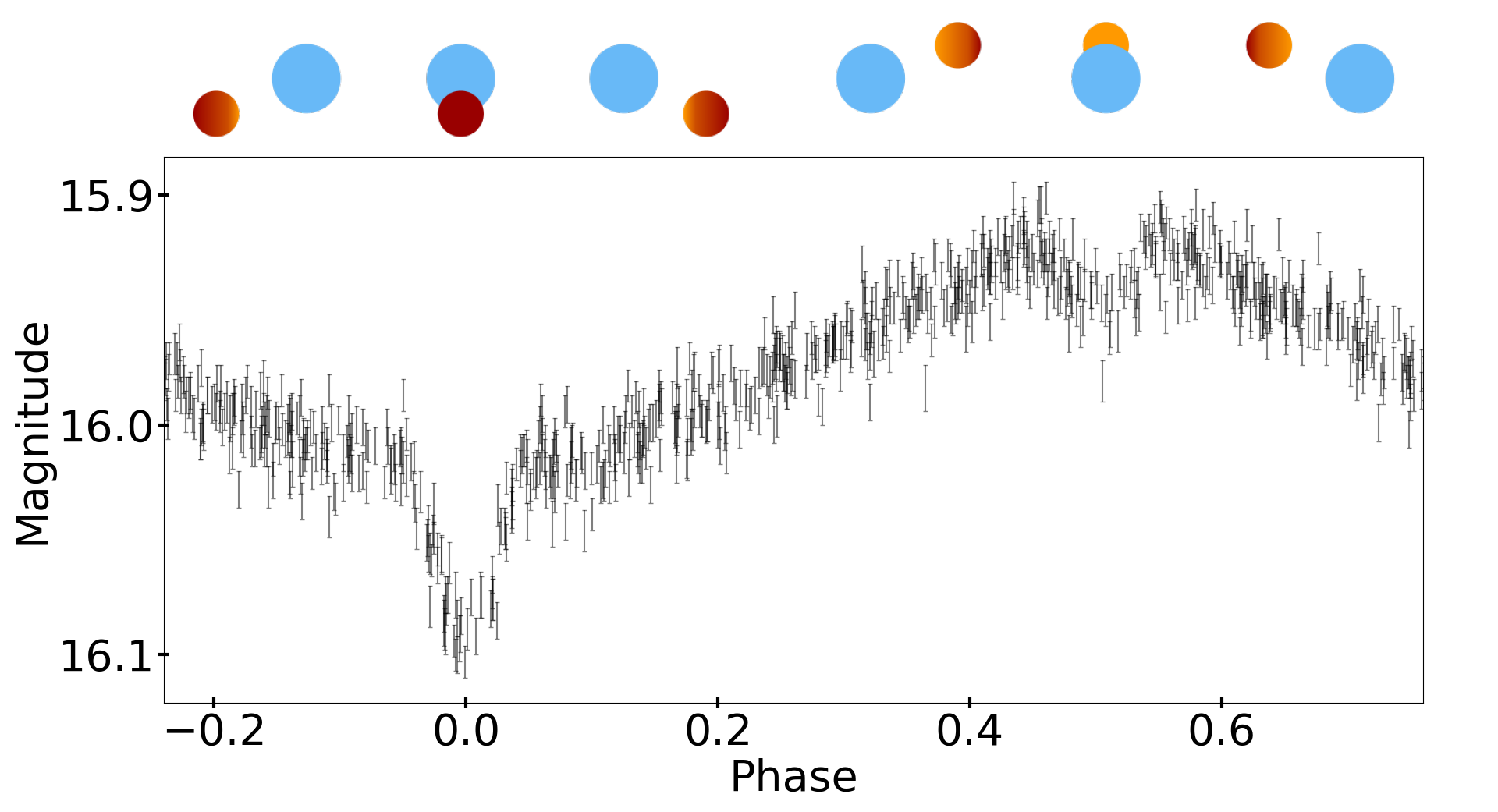}
    \caption{OGLE-III I-band light curve of nascent EB-5481 in the LMC. A diagram of the positions of the B-type MS primary (blue) and irradiated pre-MS companion (red) are displayed at the top as a function of orbital phase. }
    \label{fig:reflection}
\end{figure}

\citet{Moe2015a} discovered a new class of nascent eclipsing binaries (EBs) by analyzing the OGLE-III light curves of EBs in the LMC \citep{Gracyk2011}.  The light curves of nascent EBs exhibit reflection effects as a hot B-type main-sequence (MS) primary irradiates a close, cool, low-mass pre-MS companion (see Figure~\ref{fig:reflection}). Reflection effects greater than a few percent indicate a large temperature contrast between the two stars ($T_{\mathrm{eff,1}}$\,$\gtrsim$\,$3\,T_{\mathrm{eff,2}}$), though they must be fairly comparable in size \citep{Moe2015a}. The light curves have narrow eclipses and no ellipsoidal variability, which indicate the components are detached. Most nascent EBs are commonly found in H\,II regions. Nascent EBs with extreme mass ratios have since been discovered in star-forming regions throughout our Galaxy \citep{Jerzykiewicz2021,Pigulski2024, Naze2025}. 

\citet{Moe2015a} estimated the physical properties of nascent EBs in the LMC by fitting their OGLE-III light curves. They assumed that both components were evolving along their respective single-star evolutionary tracks. Their estimates of stellar masses, radii, temperatures, and ages were therefore uncertain and model dependent. To more accurately measure the stellar properties, multi-epoch high-resolution spectra are required. These spectra can yield the atmospheric parameters of the B-type primaries and the dynamical masses of the pre-MS secondaries. 

More accurate measurements of these nascent EBs can help constrain the formation mechanisms of close binary stars. Close binaries (a\,$<$\,10\,au) cannot form in-situ \citep{Boss1986, Bate1998, Bate2009}. Instead, they fragment and form at larger scales and migrate inwards. There are several theories as to how the inward migration occurs: disk migration \citep{Tokovinin2020}, secular evolution with a tertiary \citep{Kiseleva1998}, and dynamical interactions with other stars in their birth cluster \citep{Bate2009}. The timescale for each mechanism differs: disk migration occurs during the embedded protostellar phase while dynamical evolution with other stars occurs on much longer timescales. For example, \citet{Sana2017} and \citet{Ramirez2021} suggest that companions to massive primaries harden on timescales of $\approx$\,2\,Myr due to interactions with surrounding stars, much longer than the $\approx$\,0.2 Myr embedded protostellar phase of massive stars.  Accurately measuring the ages of nascent EBs would set limits on when the migration occurred, thereby constraining the formation mechanism.

This paper aims to measure the physical properties and ages of several nascent EBs using high-resolution spectroscopy and light-curve fitting. Section~\ref{sec:ObjSelect} discusses the selection of our objects. Section~\ref{sec:Obs} outlines the spectroscopic observations and data reduction. Section~\ref{sec:Methods} explains the methods used to measure the physical properties of both components in our nascent EBs. We summarize our main results in 
Section~\ref{sec:Summary}.

\section{Object Selection}\label{sec:ObjSelect}

We select three nascent EBs in the LMC to observe: two from~\citet{Moe2015a}, EB-2139 and EB-21641, and one from \citet{Frank2026}, EB-5481. We list each system's OGLE-III properties from~\citet{Gracyk2011} in Table~\ref{tab:OGLEproperties}. All three nascent EBs are embedded in H\,II regions within the LMC.

We fit double-Gaussian analytic models to the OGLE-III light curves as developed in \citet{Moe2015a,Moe2015b} and expanded in \citet{Mowlavi2017}. Specifically, we fit Eqns.~1 and 2 of \citet{Frank2026} to the light curves using a  Levenberg-Marquardt (LM) method to minimize the $\chi^2$ statistic. For each nascent EB, we measure the orbital period $P$ in days, epoch of primary eclipse $t_{\rm o}$ (Julian date JD $-$ 2450000), I-band magnitude immediately outside of primary eclipse $I_{\rm Out}$, depth of primary eclipse $\Delta I_1$, standard deviation of primary eclipse $\Theta_1$, the corresponding phase, depth, and standard deviation of the secondary eclipse $\Phi_2$, $\Delta I_2$, and $\Theta_2$, respectively, the full amplitude $\Delta I_{\rm Ell}$ of ellipsoidal variability (two crests per orbit peaking at quadratures $\phi$ = 0.25 and 0.75), and the full amplitude $\Delta I_{\rm Refl}$ of the reflection effect (one crest per orbit peaking at $\phi$ = 0.5). In Table~\ref{tab:OGLEfitLC}, we report the fitted parameters, number $N_{\rm I}$ of I-band data points, and reduced $\chi^2/\nu$, where the degrees of freedom $\nu$~=~$N_{\rm I}$\,$-$\,10. The reported errors for the analytic fit parameters derive from the covariance matrix of the LM technique. Each of the three light curves showcase a substantial reflection effect ($\Delta\,I_\mathrm{Refl}$\,$\geq$\,0.03 mag) with no ellipsoidal variation ($\Delta\,I_\mathrm{Ell}$\,=\,0.00 mag).

\begin{deluxetable}{lccccc}[htp]
    \setlength{\tabcolsep}{4pt}
    \tablecaption{OGLE-III Properties of Nascent EBs \label{tab:OGLEproperties}}
   \tabletypesize{\footnotesize}
    \tablehead{
        \colhead{Object} & \colhead{RA} & \colhead{Dec} & \colhead{$\langle$I$\rangle$} & \colhead{$\langle$V$\rangle-\langle$I$\rangle$} & \colhead{$P$\,(days)}
    }
    \startdata
    EB-2139 & 04:53:05.17 & -68:03:03.4 & 16.490 & $-$0.163 & 8.462520 \\
    EB-5481 & 05:01:57.57 & -70:42:07.5 & 15.978 & $-$0.136 & 2.487134 \\
    EB-21641 & 05:39:10.90 & -69:29:20.2 & 16.713 & 0.055 & 3.092410 \\
    \enddata
\end{deluxetable}

\begin{deluxetable*}{lrrrrrrrrrrcc}[htp]
    \setlength{\tabcolsep}{3pt}
    \tablecaption{Analytic Fit Parameters to OGLE-III Light Curves \label{tab:OGLEfitLC}}
    \tabletypesize{\footnotesize}
    \tablehead{
        \colhead{Object} & \colhead{$P$} & \colhead{$t_{\rm o}$} & \colhead{$I_{\rm Out}$} &
        \colhead{$\Delta I_1$} & \colhead{$\Theta_1$} &
        \colhead{$\Phi_2$} & \colhead{$\Delta I_2$} &
        \colhead{$\Theta_2$} & \colhead{$\Delta I_{\rm Ell}$} &
        \colhead{$\Delta I_{\rm Refl}$} & \colhead{$N_{\rm I}$} & \colhead{$\chi^2/\nu$} 
    }
    \startdata
    EB-2139 & 8.462510 & 3576.045 & 16.505 & 0.330 & 0.0104 & 0.501 & 0.031 & 0.009 & 0.000 & 0.030 & 477 & 1.39 \\
     ~  & $\pm$\,0.000015 & $\pm$\,0.002 & $\pm$\,0.001 & $\pm$\,0.004 & $\pm$\,0.0002 & $\pm$\,0.002 & $\pm$\,0.004 & $\pm$\,0.002 & $\pm$\,0.002 & $\pm$\,0.002 & &  \\
    EB-5481 & 2.487130 & 3561.556 & 16.021 & 0.076 & 0.0196 & 0.502 & 0.019 & 0.018 & 0.000 & 0.094 & 847 & 1.48 \\
     ~  & $\pm$\,0.000004 & $\pm$\,0.002 & $\pm$\,0.001 & $\pm$\,0.003 & $\pm$\,0.0008 & $\pm$\,0.003 & $\pm$\,0.003 & $\pm$\,0.003 & $\pm$\,0.001 & $\pm$\,0.001 & &  \\
    EB-21641 & 3.092424 & 3572.506 & 16.739 & 0.407 & 0.0194 & 0.500 & 0.073 & 0.019 & 0.000 & 0.059 & 436 & 0.81 \\
     ~  & $\pm$\,0.000005 & $\pm$\,0.002 & $\pm$\,0.003 & $\pm$\,0.007 & $\pm$\,0.0005 & $\pm$\,0.001 & $\pm$\,0.005 & $\pm$\,0.002 & $\pm$\,0.003 & $\pm$\,0.003 & &  \\
    \enddata
\end{deluxetable*}

\section{Observations and Data Reduction}\label{sec:Obs}

We obtained high-resolution multi-epoch echelle spectra of our three nascent EBs using the Magellan Inamori Kyocera Echelle (MIKE) spectrograph attached to the 6.5m Magellan-Clay Telescope at Las Campanas Observatory~\citep{Bernstein2003}. The MIKE spectrograph is a high-throughput double echelle spectrograph. We used the 1.0$''$ slit, binned 2x2, and obtained ThAr arcs for wavelength calibration at the position of each target. Each system was observed 4\,-\,7 times at differing orbital phases. We present an observation log of dates, exposure times, and seeing conditions in Table~\ref{tab:obs}. 

We reduced the data using CarPY's MIKE Pipeline, which bias-subtracted and flat-fielded the science frames, traced the apertures, and extracted the spectra from each order. We reduced and analyzed only the blue side (3,800\AA\,-\,5,000\AA), which covers many hydrogen and helium absorption features of B-type stars. We applied heliocentric corrections to the wavelength solutions of each epoch. We then normalized each order by fitting a sixth-degree polynomial to the continuum. We then co-added overlapping regions of neighboring orders. We calculate the signal-to-noise ratio (SNR) for two regions containing a featureless continuum: 4,200\,-\,4,250\,\AA\ and 4,750\,-\,4,800\AA. In Table~\ref{tab:obs}, we report these SNRs, which span 12\,-\,44 with an average value of 20. Many of the spectra contain poorly subtracted narrow Balmer emission line artifacts from the surrounding H\,II regions. The nascent EBs themselves likely do not feature narrow emission lines.

\begin{deluxetable*}{lcccccccc}[ht]
    \tablecaption{Observation Log \label{tab:obs}}
    \tablehead{
        \colhead{Object} & \colhead{UTC Date} & \colhead{UTC Time} & \colhead{Julian Date} & \colhead{Exp. Time} & \colhead{Airmass} & \colhead{Seeing} 
        & \colhead{SNR(4225\AA)} & \colhead{SNR(4775\AA)} 
    }
    \startdata
    EB-2139 & 2016-03-06 & 01:50 & 2457453.5756 & 3\,$\times$\,1200\,s & 1.4 & 0.5$''$ & 42 & 44 \\ 
            & 2016-03-07 & 03:11 & 2457454.6327 & 3\,$\times$\,1300\,s & 1.6 & 0.7$''$ & 17 & 19 \\ 
            & 2016-04-13 & 01:06 & 2457491.5459 & 3\,$\times$\,1100\,s & 1.8 & 1.3$''$ & 18 & 22 \\ 
            & 2016-04-14 & 01:09 & 2457492.5475 & 3\,$\times$\,1100\,s & 1.9 & 1.0$''$ & 21 & 26 \\ 
            & 2016-10-22 & 06:36 & 2457683.7734 & 3\,$\times$\,1000\,s & 1.3 & 1.7$''$ & 17 & 16 \\ 
            & 2017-04-03 & 02:04 & 2457846.5858 & 3\,$\times$\,1000\,s & 1.9 & 1.3$''$ & 13 & 16 \\ 
            & 2017-04-04 & 00:19 & 2457847.5134 & 3\,$\times$\,1100\,s & 1.5 & 1.0$''$ & 19 & 21 \\ 
    \hline
    EB-5481 & 2016-10-22 & 05:55 & 2457683.7446 & 3\,$\times$\,450\,s & 1.4 & 2.0$''$ & 11 & 12 \\ 
            & 2016-10-23 & 05:47 & 2457684.7408 & 3\,$\times$\,600\,s & 1.4 & 1.2$''$ & 26 & 26 \\ 
            & 2017-04-03 & 01:14 & 2457846.5512 & 3\,$\times$\,800\,s & 1.7 & 1.3$''$ & 17 & 21 \\ 
            & 2017-04-04 & 01:14 & 2457847.5513 & 3\,$\times$\,850\,s & 1.7 & 1.0$''$ & 29 & 30 \\ 
    \hline
    EB-21641 & 2016-03-06 & 04:11 & 2457453.6743 & 3\,$\times$\,1200\,s & 1.7 & 0.5$''$ & 23 & 31 \\ 
             & 2016-03-07 & 00:41 & 2457454.5286 & 3\,$\times$\,1300\,s & 1.4 & 0.8$''$ & 19 & 20 \\ 
             & 2016-04-13 & 00:06 & 2457491.5044 & 3\,$\times$\,1100\,s & 1.5 & 1.4$''$ & 11 & 12 \\ 
             & 2016-04-14 & 02:13 & 2457492.5925 & 3\,$\times$\,1260\,s & 1.0 & 1.1$''$ & 13 & 19 \\ 
             & 2016-10-23 & 06:34 & 2457684.7737 & 3\,$\times$\,1100\,s & 1.3 & 1.1$''$ & 23 & 24 \\ 
             & 2017-04-02 & 02:21 & 2457845.5972 & 3\,$\times$\,1000\,s & 1.8 & 1.4$''$ & 12 & 13 \\ 
             & 2017-04-03 & 03:03 & 2457846.6271 & 3\,$\times$\,1150\,s & 2.0 & 1.3$''$ & 10 & 14 \\ 
    \enddata
\end{deluxetable*}

\section{Methods} \label{sec:Methods}
\subsection{Properties of B-dwarf Primaries} \label{sec:PrimParams}

\subsubsection{Atmospheric Parameters} \label{sec:AtmParams}

In each spectrum, only absorption features from the luminous B-dwarf primary are visible. For that reason, we choose to fit TLUSTY BSTAR2006 models with LMC metallicity (Z\,=\,0.5\,Z$_{\odot}$) to the normalized spectra. We downloaded TLUSTY models with effective temperatures $T_{\rm eff}$ = 15,000\,K\,-\,30,000\,K and surface gravities \logg = 2.0\,-\,4.75. We smooth each model with a Gaussian kernel of varying projected rotation velocity, spanning \vsini\ = 30\,-\,300\,\ks\ in increments of 5\,\ks. Additionally, we shift each model in steps of 1\,km\,$\mathrm{s}^{-1}$ in order to determine the radial velocity $v_{\rm r}$ for each epoch.

\begin{figure*}
    \centering
    \includegraphics[width=\linewidth]{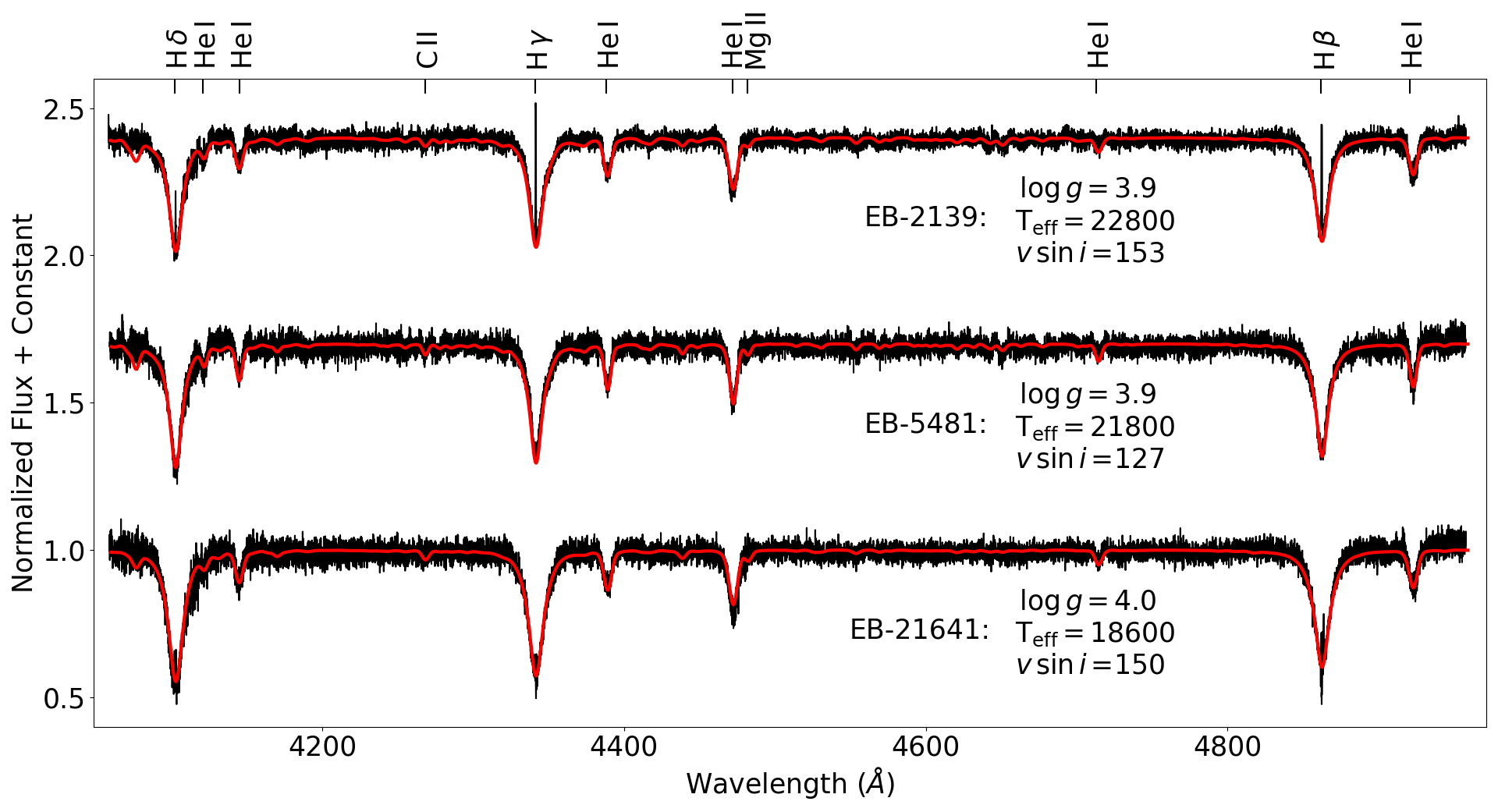}
    \caption{Stacked spectrum of each nascent EB (black) with corresponding best-fit TLUSTY model (red). Notable spectral lines are indicated at the top, and the adopted parameters are shown below each spectrum.}
    \label{fig:multiSpec}
\end{figure*}

We minimize the \cs statistic between our 4D grid of TLUSTY models and each spectrum from each epoch. We stack the spectra of each object by shifting them according to their measured radial velocities $v_{\rm r}$. We also fit the TLUSTY models to our stacked spectra. We display the fits to the stacked spectra of each object in Figure~\ref{fig:multiSpec}. We report the fitted parameters for each spectrum in Table~\ref{tab:AtmFitResults}. For $T_{\rm eff}$ and $v$\,sin\,$i$, we adopt values and uncertainties based on the averages and standard deviations of the multiple epochs. For log\,$g$, we adopt an uncertainty of 0.2~dex, slightly smaller than the 0.25~dex spacings between TLUSTY models. 

\begin{deluxetable*}{lcccccc}[ht]
    \setlength{\tabcolsep}{3pt}
    \tablecaption{Atmospheric Model Fits of B-dwarf Primaries \label{tab:AtmFitResults}}
    \tablehead{
        \colhead{Object} & \colhead{Julian Date} & \colhead{$T_{\rm eff, 1}$ (K)} & \colhead{log\,$g$} & \colhead{$v$\,sin\,$i$ (km\,s$^{-1}$)} &  \colhead{$v_{\rm r}$ (km\,s$^{-1}$)} & \colhead{$\chi^2/\nu$}
    }
    \startdata
    EB-2139 & 2457453.5756 & 22000 & 3.75 & 160 & 289 & 1.16 \\ 
            & 2457454.6327 & 24000 & 4.00 & 145 & 286 & 1.11 \\ 
            & 2457491.5459 & 21000 & 3.75 & 160 & 304 & 1.15 \\ 
            & 2457492.5475 & 23000 & 4.00 & 155 & 302 & 1.20 \\ 
            & 2457683.7734 & 22000 & 4.00 & 145 & 286 & 1.08 \\ 
            & 2457846.5858 & 24000 & 4.00 & 155 & 309 & 1.16 \\ 
            & 2457847.5134 & 24000 & 3.75 & 145 & 310 & 1.11 \\ 
            & Stacked & 22000 & 3.75 & 155 & - & 2.09 \\
            & Adopted & 22800\,$\pm$\,1200 & 3.9\,$\pm$\,0.2 & 153\,$\pm$\,7 & - & - \\
    \hline
    EB-5481 & 2457683.7446 & 23000 & 4.00 & 135 & 236 & 1.25 \\ 
            & 2457684.7408 & 23000 & 4.00 & 120 & 283 & 1.12 \\ 
            & 2457846.5512 & 20000 & 3.75 & 125 & 296 & 1.10 \\ 
            & 2457847.5513 & 22000 & 3.75 & 125 & 222 & 1.16 \\ 
            & Stacked & 21000 & 3.75 & 130 & - & 1.20 \\
            & Adopted & 21800\,$\pm$\,1400 & 3.9\,$\pm$\,0.2 & 127\,$\pm$\,7 & - & - \\
    \hline
    EB-21641 & 2457453.6743 & 19000 & 4.00 & 140 & 248 & 1.18 \\ 
             & 2457454.5286 & 20000 & 4.00 & 150 & 244 & 1.03 \\ 
             & 2457491.5044 & 18000 & 4.00 & 165 & 264 & 1.11 \\ 
             & 2457492.5925 & 19000 & 4.00 & 155 & 298 & 1.10 \\ 
             & 2457684.7737 & 18000 & 4.00 & 140 & 293 & 1.13 \\ 
             & 2457845.5972 & 18000 & 3.75 & 155 & 289 & 0.99 \\ 
             & 2457846.6271 & 18000 & 4.00 & 150 & 267 & 1.08 \\ 
             & Stacked & 19000 & 4.00 & 145 & - & 1.71 \\
             & Adopted & 18600\,$\pm$\,800 & 4.0\,$\pm$\,0.2 & 150\,$\pm$\,9 & - & - \\
    \enddata
\end{deluxetable*}

\subsubsection{SED Fitting of B-dwarf Primaries}\label{sec:SEDFitting}

\begin{figure}
    \centering
    \begin{subfigure}{0.9\linewidth}
        \includegraphics[width=\linewidth]{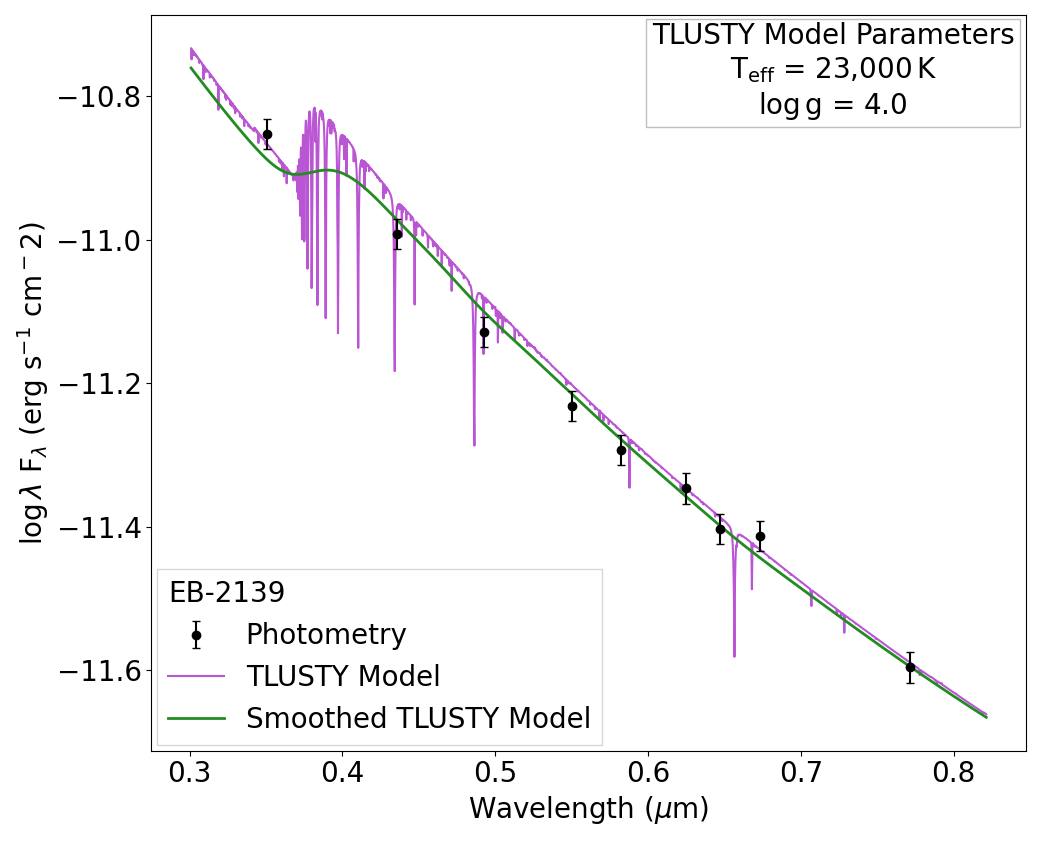}
        \caption{EB-2139}
        \label{fig:EB2139SED}
    \end{subfigure}
    \hfill
    \begin{subfigure}{0.9\linewidth}
        \includegraphics[width=\linewidth]{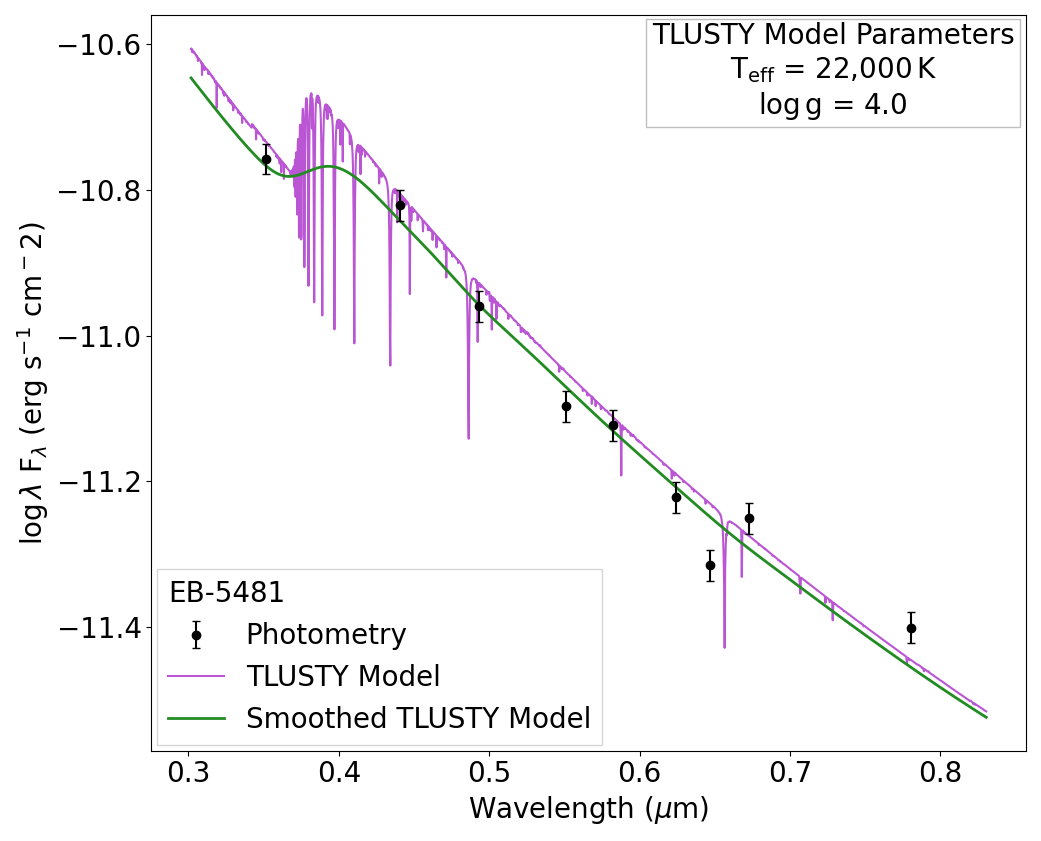}
        \caption{EB-5481}
        \label{fig:EB5481SED}
    \end{subfigure}
    \hfill
    \begin{subfigure}{0.9\linewidth}
        \includegraphics[width=\linewidth]{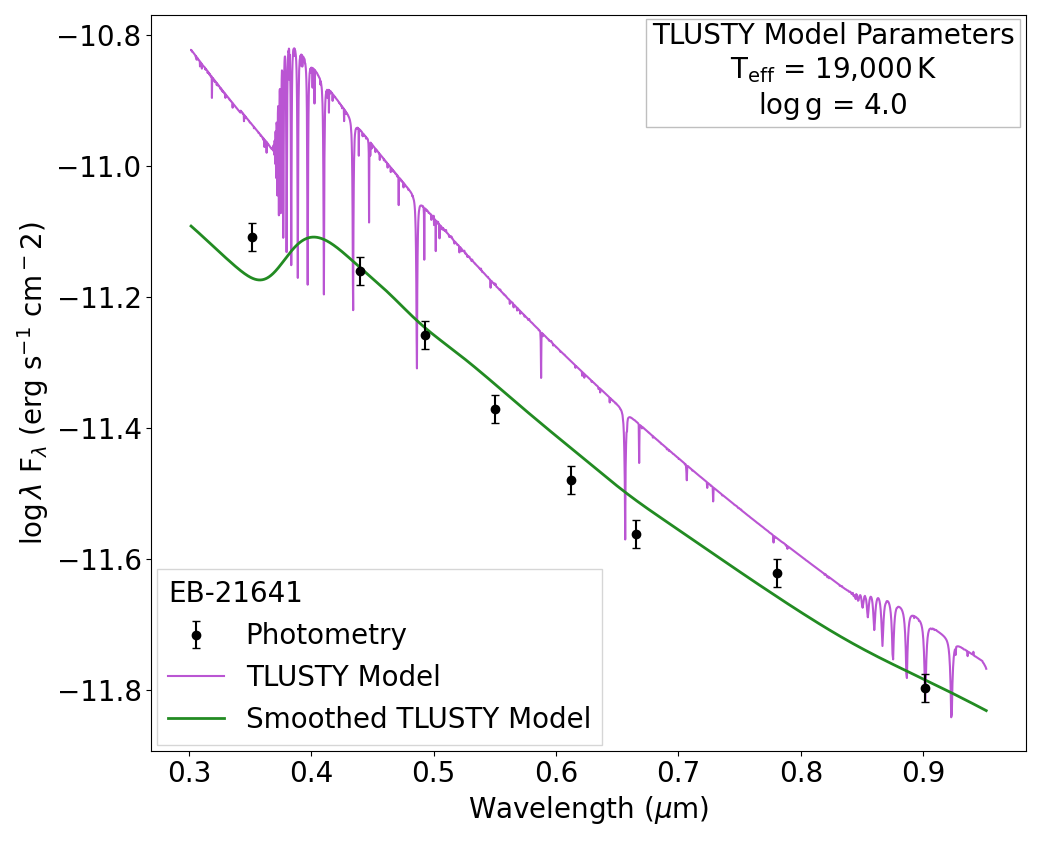}
        \caption{EB-21641}
        \label{fig:EB21641SED}
    \end{subfigure}
    
    \caption{TLUSTY model atmospheres (purple) and smoothed, dust-reddened SED fits (green) to broadband photometry (black).}
    \label{fig:SEDs}
\end{figure}

We next fit spectral energy distributions (SEDs) to broadband photometry of our nascent EBs in order to measure the luminosities of the B-dwarf primaries. We collect data from Vizier's Photometry Viewer\footnote{http://vizier.cds.unistra.fr/vizier/sed/}. To estimate the brightness of the EB just outside of primary eclipse, we adopt the 20th percentile in photometry for each passband and assume 5\% uncertainties in the fluxes (see also \citealt{Frank2026}). For the unreddened SEDs, we select the TLUSTY model closest to our adopted $T_{\rm eff}$ and log\,$g$ based on the atmospheric fits. We smooth the SEDs by a Gaussian kernel of 200\AA\ to roughly match the widths of the photometric passbands. We adopt a distance $d$ = 49.6 kpc to the LMC \citep[distance modulus $\mu$ = 18.48;][]{Pietrzynski2019}. We also adopt a \citet{Fitzpatrick1999} dust-reddening law with $R_{\rm V}$ = 3.4, appropriate for LMC dust. Our models have two free parameters: luminosity $L_1$ of the B-dwarf primary and visual dust extinction A$_{\rm V}$. We implement an LM technique to minimize the $\chi^2$ statistic between our models and photometric data. We report the best-fit parameters in Table~\ref{tab:SEDfit} and present the best-fit SEDs in Figure~\ref{fig:SEDs}. The reported errors in $L_1$ and A$_{\rm V}$ derive from the covariance matrix of the LM technique. 

Given our adopted dust-reddening law, we compute the I-band dust extinction A$_{\rm I}$ and dust reddening E(V$-$I). For comparison, we also report in Table~\ref{tab:SEDfit} the dust reddening E(V$-$I)$_{\rm OGLE}$ inferred from OGLE-IV red clump stars at the coordinates of our nascent EBs \citep{Skowron2021}. Our measured dust reddenings are consistent with those inferred from OGLE-IV red clump stars. EB-2139 and EB-5481 experience only slight dust reddening while EB-21641 is highly embedded. We compute the absolute I-band magnitudes M$_{\rm I}$ = I$_{\rm Out}$\,$-$\,$\mu$\,$-$A$_{\rm I}$. Finally, we estimate the radii $R_1$ of the B-dwarf primaries based on the measured $T_{\rm eff,1}$ from the atmospheric fits, $L_1$ from the SED fits, and the Stefan-Boltzmann law. 

\begin{deluxetable*}{lcccccccccc}[t]
    \tablecaption{SED Fit Parameters of B-dwarf Primaries \label{tab:SEDfit}}
    \tablehead{
        \colhead{Object} & \colhead{A$_{\rm V}$ (mag)} & \colhead{$L_1$ (\Lsun)} & \colhead{$N_{\rm p}$} & \colhead{$\chi^2/\nu$} & \colhead{A$_{\rm I}$ (mag)} & \colhead{E(V$-$I)} &  \colhead{E(V$-$I)$_{\rm OGLE}$} & \colhead{M$_{\rm I}$ (mag)} & \colhead{$R_1$ (\Rsun)} & \colhead{M$_1$(\Msun)}
    }
    \startdata
    EB-2139 & 0.04\,$\pm$\,0.07 & 5300\,$\pm$\,400 & 9 & 1.20 & 0.03\,$\pm$\,0.04 & 0.02\,$\pm$\,0.08 & 0.13\,$\pm$\,0.05 & $-$2.00\,$\pm$\,0.04 & 4.6\,$\pm$\,0.4 & 8.3\,$\pm$\,0.7 \\
    \hline
    EB-5481 & 0.07\,$\pm$\,0.07 & 6800\,$\pm$\,400 & 9 & 3.32 & 0.04\,$\pm$\,0.04 & 0.03\,$\pm$\,0.08 & 0.09\,$\pm$\,0.05 & $-$2.50\,$\pm$\,0.04 & 5.7\,$\pm$\,0.5 & 8.5\,$\pm$\,0.8 \\
    \hline
    EB-21641 & 0.40\,$\pm$\,0.06 & 3600\,$\pm$\,200 & 8 & 6.49 & 0.22\,$\pm$\,0.04 & 0.18\,$\pm$\,0.07 & 0.34\,$\pm$\,0.18 & $-$1.96\,$\pm$\,0.04 & 5.6\,$\pm$\,0.4 & 6.8\,$\pm$\,0.6 \\
    \enddata
\end{deluxetable*}

\subsubsection{Masses of B-dwarf Primaries} \label{sec:massB}

\begin{figure}
    \centering
    \includegraphics[width=\linewidth]{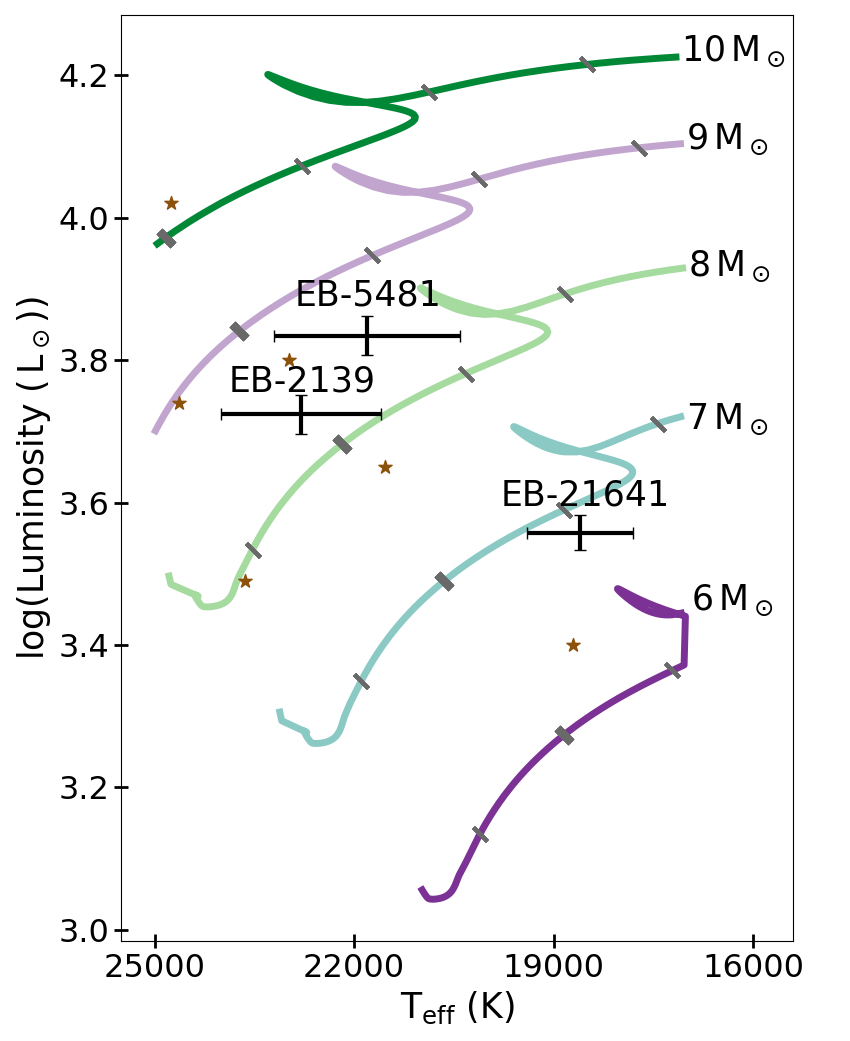}
    \caption{Locations of our three B-dwarf primaries (black crosses) alongside the primaries in Galactic nascent EBs from \citet{Naze2025} (brown stars) on an HRD. We overlay MIST evolutionary tracks, plotting surface gravity tick marks in intervals of 0.25 dex (thick tick mark at log\,$g$ = 4.0).}
    \label{fig:HRD1}
\end{figure}

We next estimate the masses $M_{\mathrm{1}}$ of the primary B-dwarfs based on their locations on the Hertzsprung-Russell diagram (HRD). Our nascent EBs are well detached (see Section~\ref{sec:ObjSelect}) and so both components are evolving along their respective single-star evolutionary tracks. In Figure~\ref{fig:HRD1}, we present MIST MS evolutionary tracks for $M_1$ = 6\,-\,10\,\Msun\ primaries at LMC metallicity Z\,=\,0.5\,Z$_{\odot}$ \citep{Choi2016}. We indicate tick marks with respect to surface gravity in intervals of 0.25 dex (thick tick mark at \logg\,=\,4.0). We plot our three mid-B primaries in Figure~\ref{fig:HRD1} alongside the six mid-B primaries in Galactic nascent EBs from \citet{Naze2025}. Despite being nascent EBs, the primaries appear to be slightly evolved along the MS, approaching the terminal-age MS. Nonetheless, the surface gravities inferred from the HRD (log\,$g$ = 3.75 - 4.00) are consistent with our measurements based on the atmospheric fits. We interpolate the evolutionary tracks to estimate the masses $M_1$ of our B-dwarf primaries. We add a systematic error of 0.4~\Msun\ in quadrature with the uncertainties derived from propagating the errors in $T_{\rm eff,1}$ and $L_1$. Our primary masses span $M_1$ = 6.8\,-\,8.5\,\Msun.

\subsection{Dynamical Masses of the Pre-MS Secondaries}
\label{sec:RVcurves}

\subsubsection{Radial Velocities} \label{sec:adoptedRadVel}

We now measure the masses of the pre-MS secondaries based on the radial velocity curves of the mid-B dwarf primaries. We already obtained radial velocities for each epoch from the atmospheric fits described in Section~\ref{sec:AtmParams}. We report the radial velocities and uncertainties from the $\chi^2$ minimization of the atmospheric fits in Table~\ref{tab:Rv}. Additionally, we fit Gaussian profiles to two deep He lines (4387\,\AA\ and 4923\,\AA) and the Doppler cores of three Balmer hydrogen lines (H\,$\beta$, H\,$\gamma$, and H\,$\delta$). We exclude the shallower or blended features, which have large uncertainties. We report the radial velocities and uncertainties for the five lines in Table~\ref{tab:Rv}. We also compute a weighted average of the five lines. The radial velocities inferred from the atmospheric fits and weighted average of the five lines are sometimes slightly discrepant with each other. We therefore adopt a weighted average of the two methods and adopt uncertainties that bracket the 84th percentile (1$\sigma$) of the two different techniques.

\begin{deluxetable*}{lccrrrrrrrr}[ht]
    \setlength{\tabcolsep}{4pt}
    \tablecaption{Radial Velocities (km s$^{-1}$) \label{tab:Rv}}
    \tablehead{
        \colhead{Object} & \colhead{Julian} & \colhead{Phase} & \colhead{H$\delta$} & \colhead{H$\gamma$} & \colhead{H$\beta$} & \colhead{He\,I} & \colhead{He\,I} & \colhead{Weighted} & \colhead{$\chi ^2$} & \colhead{Adopted} \\
        \colhead{} & \colhead{Date} & \colhead{} & \colhead{} & \colhead{} & \colhead{} & \colhead{4387} & \colhead{4923} & \colhead{Average} & \colhead{Method} & \colhead{} 
    }
    \startdata
    EB-2139 & 2457453.5756 & 0.201 & 282.1 & 292.5 & 276.0 & 295.3 & 276.0 & 285.2 & 289.0 & 287 \\
            &  & & \,$\pm$\,2.7 & \,$\pm$\,2.3 & \,$\pm$\,2.8 & \,$\pm$\,4.4 & \,$\pm$\,5.9 & \,$\pm$\,1.4 & \,$\pm$\,1.1 & \,$\pm$\,3 \\ 
            & 2457454.6327 & 0.326 & 272.5 & 311.1 & 274.6 & 287.8 & 273.4 & 288.2 & 286.0 & 287 \\
            &  & & \,$\pm$\,8.4 & \,$\pm$\,5.6 & \,$\pm$\,5.8 & \,$\pm$\,8.5 & \,$\pm$\,12.1 & \,$\pm$\,3.2 & \,$\pm$\,2.5 & \,$\pm$\,4 \\  
            & 2457491.5459 & 0.688 & 294.8 & 312.1 & 296.2 & 303.6 & 280.9 & 299.1 & 304.0 & 302 \\
            &  & & \,$\pm$\,6.5 & \,$\pm$\,5.6 & \,$\pm$\,4.6 & \,$\pm$\,10.2 & \,$\pm$\,9.0 & \,$\pm$\,2.8 & \,$\pm$\,2.2 & \,$\pm$\,4 \\ 
            & 2457492.5475 & 0.806 & 307.7 & 301.4 & 293.9 & 310.3 & 271.6 & 299.5 & 302.0 & 301 \\
            &  & & \,$\pm$\,5.3 & \,$\pm$\,4.4 & \,$\pm$\,4.3 & \,$\pm$\,7.5 & \,$\pm$\,10.1 & \,$\pm$\,2.4 & \,$\pm$\,1.9 & \,$\pm$\,3 \\ 
            & 2457683.7734 & 0.403 & 283.0 & 285.0 & 287.8 & 301.4 & 265.5 & 285.9 & 286.0 & 286 \\
            &  & & \,$\pm$\,6.7 & \,$\pm$\,5.0 & \,$\pm$\,6.1 & \,$\pm$\,9.7 & \,$\pm$\,13.2 & \,$\pm$\,3.1 & \,$\pm$\,0.3 & \,$\pm$\,3 \\ 
            & 2457846.5858 & 0.642 & 312.1 & 328.6 & 290.6 & 305.8 & 307.3 & 310.2 & 309.0 & 309 \\
            &  & & \,$\pm$\,10.5 & \,$\pm$\,6.7 & \,$\pm$\,7.0 & \,$\pm$\,16.8 & \,$\pm$\,17.7 & \,$\pm$\,4.1 & \,$\pm$\,2.6 & \,$\pm$\,4 \\ 
            & 2457847.5134 & 0.752 & 302.1 & 305.4 & 295.3 & 299.6 & 287.6 & 300.5 & 310.0 & 305 \\
            &  & & \,$\pm$\,5.9 & \,$\pm$\,5.2 & \,$\pm$\,6.4 & \,$\pm$\,8.3 & \,$\pm$\,11.8 & \,$\pm$\,3.0 & \,$\pm$\,2.2 & \,$\pm$\,6 \\ 
    \hline
    EB-5481 & 2457683.7446 & 0.408 & 253.5 & 226.4 & 213.2 & 240.8 & 214.2 & 229.3 & 236.0 & 233 \\
            &  & & \,$\pm$\,9.1 & \,$\pm$\,7.4 & \,$\pm$\,7.5 & \,$\pm$\,11.2 & \,$\pm$\,18.0 & \,$\pm$\,4.1 & \,$\pm$\,2.7 & \,$\pm$\,7 \\ 
            & 2457684.7408 & 0.809 & 286.5 & 282.7 & 274.5 & 282.9 & 261.5 & 280.2 & 283.0 & 282 \\
            &  & & \,$\pm$\,3.9 & \,$\pm$\,3.2 & \,$\pm$\,3.5 & \,$\pm$\,4.6 & \,$\pm$\,7.6 & \,$\pm$\,1.8 & \,$\pm$\,1.5 & \,$\pm$\,3 \\  
            & 2457846.5512 & 0.868 & 294.2 & 298.8 & 274.4 & 298.3 & 303.0 & 290.8 & 296.0 & 293 \\
            &  & & \,$\pm$\,5.8 & \,$\pm$\,4.3 & \,$\pm$\,4.5 & \,$\pm$\,7.5 & \,$\pm$\,9.6 & \,$\pm$\,2.5 & \,$\pm$\,1.9 & \,$\pm$\,4 \\ 
            & 2457847.5513 & 0.270 & 227.6 & 219.6 & 210.4 & 210.6 & 216.2 & 217.4 & 222.0 & 220 \\
            &  & & \,$\pm$\,3.9 & \,$\pm$\,3.1 & \,$\pm$\,3.3 & \,$\pm$\,5.6 & \,$\pm$\,7.2 & \,$\pm$\,1.8 & \,$\pm$\,1.3 & \,$\pm$\,4 \\ 
    \hline
    EB-21641 & 2457453.6743 & 0.057 & 254.6 & 248.9 & 233.0 & 248.5 & 230.5 & 243.3 & 248.0 & 245 \\
             &  & & \,$\pm$\,4.7 & \,$\pm$\,3.2 & \,$\pm$\,3.3 & \,$\pm$\,6.2 & \,$\pm$\,7.4 & \,$\pm$\,1.9 & \,$\pm$\,1.5 & \,$\pm$\,4 \\ 
             & 2457454.5286 & 0.333 & 232.3 & 250.0 & 235.2 & 218.6 & 228.7 & 238.1 & 244.0 & 241 \\
             &  & & \,$\pm$\,5.7 & \,$\pm$\,4.3 & \,$\pm$\,4.9 & \,$\pm$\,8.4 & \,$\pm$\,11.2 & \,$\pm$\,2.6 & \,$\pm$\,2.0 & \,$\pm$\,5 \\  
             & 2457491.5044 & 0.290 & 290.6 & 268.8 & 259.8 & 302.0 & 226.3 & 271.2 & 264.0 & 267 \\
             &  & & \,$\pm$\,9.9 & \,$\pm$\,7.3 & \,$\pm$\,8.2 & \,$\pm$\,18.6 & \,$\pm$\,24.8 & \,$\pm$\,4.6 & \,$\pm$\,2.7 & \,$\pm$\,8 \\ 
             & 2457492.5925 & 0.642 & 290.2 & 275.7 & 290.7 & 335.0 & 292.1 & 289.5 & 298.0 & 294 \\
             &  & & \,$\pm$\,7.5 & \,$\pm$\,6.0 & \,$\pm$\,6.1 & \,$\pm$\,12.1 & \,$\pm$\,11.1 & \,$\pm$\,3.4 & \,$\pm$\,2.3 & \,$\pm$\,7 \\ 
             & 2457684.7737 & 0.788 & 288.3 & 292.1 & 291.9 & 297.1 & 299.3 & 291.8 & 293.0 & 292 \\
             &  & & \,$\pm$\,4.1 & \,$\pm$\,3.4 & \,$\pm$\,4.5 & \,$\pm$\,6.9 & \,$\pm$\,10.7 & \,$\pm$\,2.1 & \,$\pm$\,1.6 & \,$\pm$\,3 \\ 
             & 2457845.5972 & 0.793 & 281.8 & 287.4 & 291.9 & 276.7 & 271.2 & 285.7 & 289.0 & 287 \\
             &  & & \,$\pm$\,9.5 & \,$\pm$\,6.6 & \,$\pm$\,8.0 & \,$\pm$\,17.9 & \,$\pm$\,15.1 & \,$\pm$\,4.2 & \,$\pm$\,2.7 & \,$\pm$\,4 \\ 
             & 2457846.6271 & 0.126 & 289.9 & 260.5 & 249.6 & 220.0 & 251.4 & 256.4 & 267.0 & 261 \\
             &  & & \,$\pm$\,13.1 & \,$\pm$\,6.9 & \,$\pm$\,6.9 & \,$\pm$\,17.9 & \,$\pm$\,13.6 & \,$\pm$\,4.2 & \,$\pm$\,3.1 & \,$\pm$\,9 \\ 
    \enddata
\end{deluxetable*}

\subsubsection{Radial Velocity Curves} \label{sec:radVelCurves}

We next fit the radial velocity curve of each system. Our nascent EBs are tidally circularized (eccentricity $e$ = 0.00) based on the measured secondary eclipse phase ($\Phi_2$~=~0.5) and eclipse widths ($\Theta_1$ $\approx$ $\Theta_2$) from our analytic fits (Section~\ref{sec:ObjSelect} and Table~\ref{tab:OGLEfitLC}). Thus, we use a sinusoidal curve to fit the adopted radial velocities in Table~\ref{tab:Rv}. We adopt the period $P$ and epoch of primary eclipse $t_{\rm o}$ from the analytic fits to calculate the phase of each epoch. We then use a \cs minimization technique to fit the systemic velocity $\gamma$ and velocity semi-amplitude $K_\mathrm{1}$ for each object. We present our best-fit radial velocities curves in Figures~\ref{fig:rvlc2139}, ~\ref{fig:rvlc5481}, and~\ref{fig:rvlc21641} and the results of the fits in Table~\ref{tab:rvFits}. We utilize the measured parameters $K_\mathrm{1}$, $P$, and $M_\mathrm{1}$ to calculate $M_\mathrm{2}$\,sin\,$i$ using the standard relation:

\begin{equation}\label{eqn:k1}
    K_1 = \left(\frac{2\pi{\rm G}}{P}\right)^{\frac{1}{3}} 
    \frac{M_2}{M_1}
    \left(M_1+M_2\right)^{\frac{1}{3}}\frac{{\rm sin}\,i}{\sqrt{1-e^2}} .
\end{equation}

\noindent We report the results in Table~\ref{tab:rvFits}.

\begin{deluxetable}{lccccc}[ht]
    \setlength{\tabcolsep}{2pt}
    \tablecaption{Radial Velocity Curve Fit Parameters \label{tab:rvFits}}
    \tablehead{
        \colhead{Object} & \colhead{$\gamma$\,(km\,s$^{-1}$)} & \colhead{$K_\mathrm{1}$\,(km\,s$^{-1}$)} & \colhead{$N_{\rm e}$} & \colhead{$\chi^2/\nu$} & \colhead{$M_2$\,sin\,$i$ (\Msun)}
    }
    \startdata
    EB-2139 & 295\,$\pm$\,2  & 10\,$\pm$\,2 & 7 & 1.06 & 0.39\,$\pm$\,0.07 \\
    \hline
    EB-5481 & 255\,$\pm$\,2  & 35\,$\pm$\,2 & 4 & 6.46 & 0.90\,$\pm$\,0.10 \\
    \hline
    EB-21641 & 265\,$\pm$\,2  & 25\,$\pm$\,2 & 7 & 4.81 & 0.60\,$\pm$\,0.07 \\ 
    \enddata
\end{deluxetable}

\subsection{Light Curve Fitting}
\label{sec:LCfitting}

\subsubsection{PHOEBE Models}\label{sec:PHOEBEModels}

We next utilize the PHysics of Eclipsing BinariEs (PHOEBE) software\footnote{https://phoebe-project.org/} \citep{Prsa2005} to fit models to the OGLE-III I-band light curves of our nascent EBs. For each system, we adopt $P$ and $t_{\rm o}$ determined from the analytic fits (Table~\ref{tab:OGLEfitLC}), primary effective temperatures $T_{\rm eff,1}$ based on the TLUSTY atmospheric fits to the spectra (Table~\ref{tab:AtmFitResults}), primary masses $M_1$ inferred from their locations on the HRDs (Table~\ref{tab:SEDfit}), and $M_2$\,sin\,$i$ measured from the radial velocity curves (Table~\ref{tab:rvFits}). We set the eccentricities to zero because our nascent EBs are tidally circularized (see Section~\ref{sec:ObjSelect}).  We adopt a logarithmic limb-darkening law for both components. For the hot B-type primaries, we interpolate the limb-darkening coefficients from Table~1 of \citet{VanHamme1993} with respect to log\,$g$ and $T_{\rm eff}$. For the cool pre-MS secondaries, we adopt the limb-darkening coefficients based on model atmospheres internal to PHOEBE. Finally, we set the primary albedo to unity, i.e., $A_1$ = 1.0, which is appropriate for the hot radiative atmospheres of our mid-B primaries. We fit five model parameters: inclination $i$, primary radius $R_\mathrm{1}$, secondary radius $R_\mathrm{2}$, secondary effective temperature $T_\mathrm{eff,2}$, and secondary albedo $A_\mathrm{2}$. Although we already estimated $R_\mathrm{1}$ from fitting the SEDs (see Section~\ref{sec:SEDFitting}), we elect to fit this parameter more precisely from the light curves. The primary radii estimated from the SEDs are relatively uncertain ($\delta R_1$/$R_1$ $\approx$ 8\%), whereas PHOEBE can generally measure radii to an accuracy of a few percent. We thus keep $R_1$ as a free parameter in our light curve models. 

We create a large 5D grid of PHOEBE models that span our five model parameters. Typical spacings are 0.05\,\Rsun\ in both $R_\mathrm{1}$ and $R_\mathrm{2}$, 200\,K in $T_\mathrm{eff,2}$, 5\% in $A_\mathrm{2}$, and 0.1$^{\circ}$ in $i$. We compute probability density functions (PDFs) of each parameter to ensure our model grids encompass the best-fit solution. We then minimize the \cs statistic between the PHOEBE models and the OGLE-III light curves. As done in \citet{Frank2026}, we fit the light curves three successive times with scipy's curve\_fit package \citep{SciPy}. Starting with the expected values from the PDFs as the initial guess, we apply a Trust Region Reflective (TRF) algorithm to fit the parameters. Using the best-fit values as the next initial guess, we repeat the TRF process, but now allow outlier data points that exceed 5$\sigma$ to be clipped. In the present study, no data points were actually clipped. Finally, using the updated best-fit values as the final initial guess, we minimize the \cs statistic with a LM technique. 

\subsubsection{Results of Light Curve Fits}\label{sec:ResultsLCFit}

\begin{figure*}[hbtp]
    \centering
    \includegraphics[width=\linewidth]{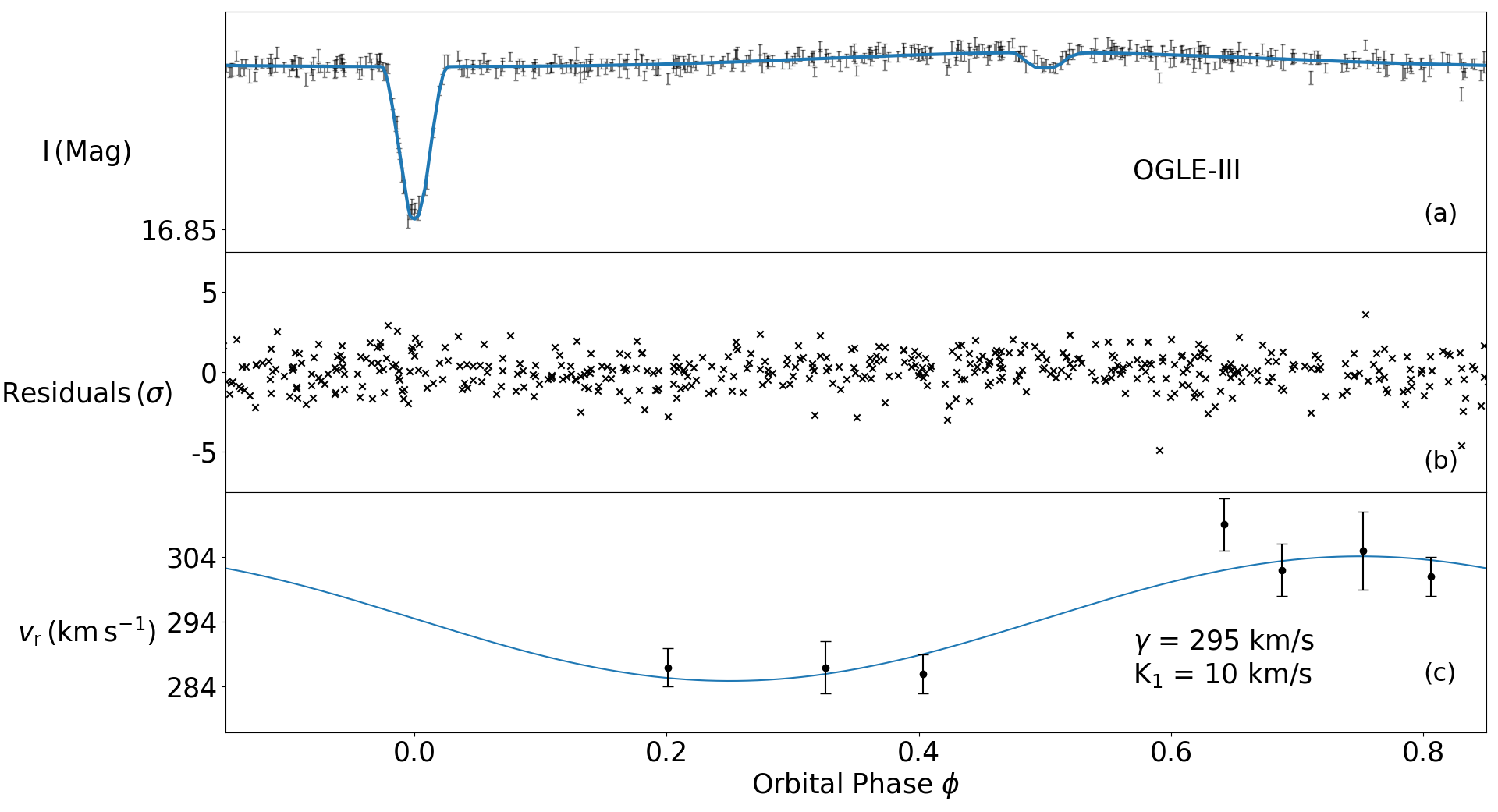}
    \caption{EB-2139 light curves and radial velocity fits. Panel \textit{a} shows the light curve from OGLE-III (black) and the corresponding best-fit PHOEBE model (blue). Panel \textit{b} shows the residuals between the OGLE-III light curve and PHOEBE model. Panel \textit{c} shows the adopted radial velocities (black) and corresponding fit (blue). }
    \label{fig:rvlc2139}
\end{figure*}

\begin{figure*}[htbp]
    \centering
    \includegraphics[width=\linewidth]{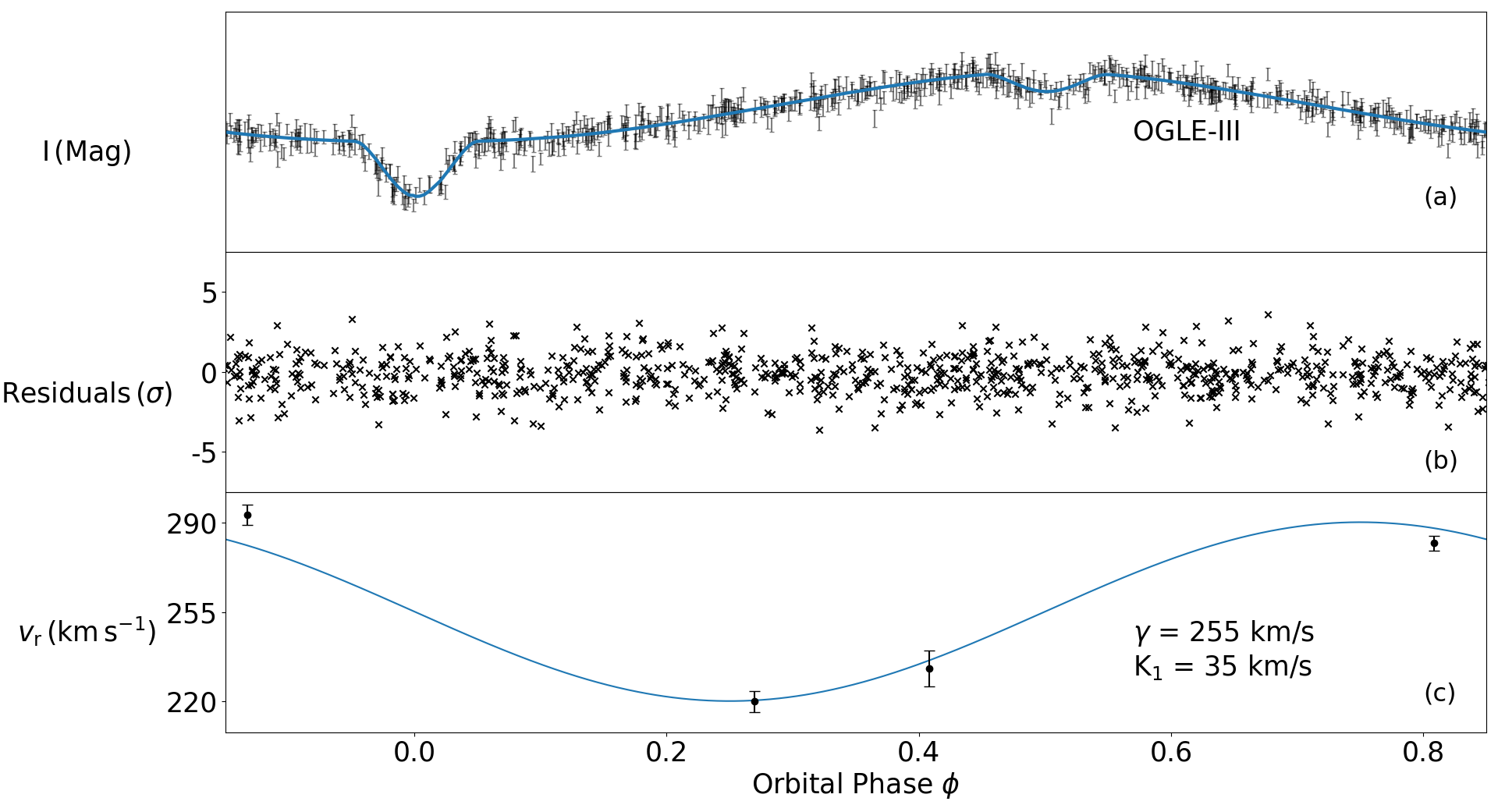}
    \caption{Same as Figure~\ref{fig:rvlc2139} but for EB-5481}
    \label{fig:rvlc5481}
\end{figure*}

\begin{figure*}[htp]
    \centering
    \includegraphics[width=\linewidth]{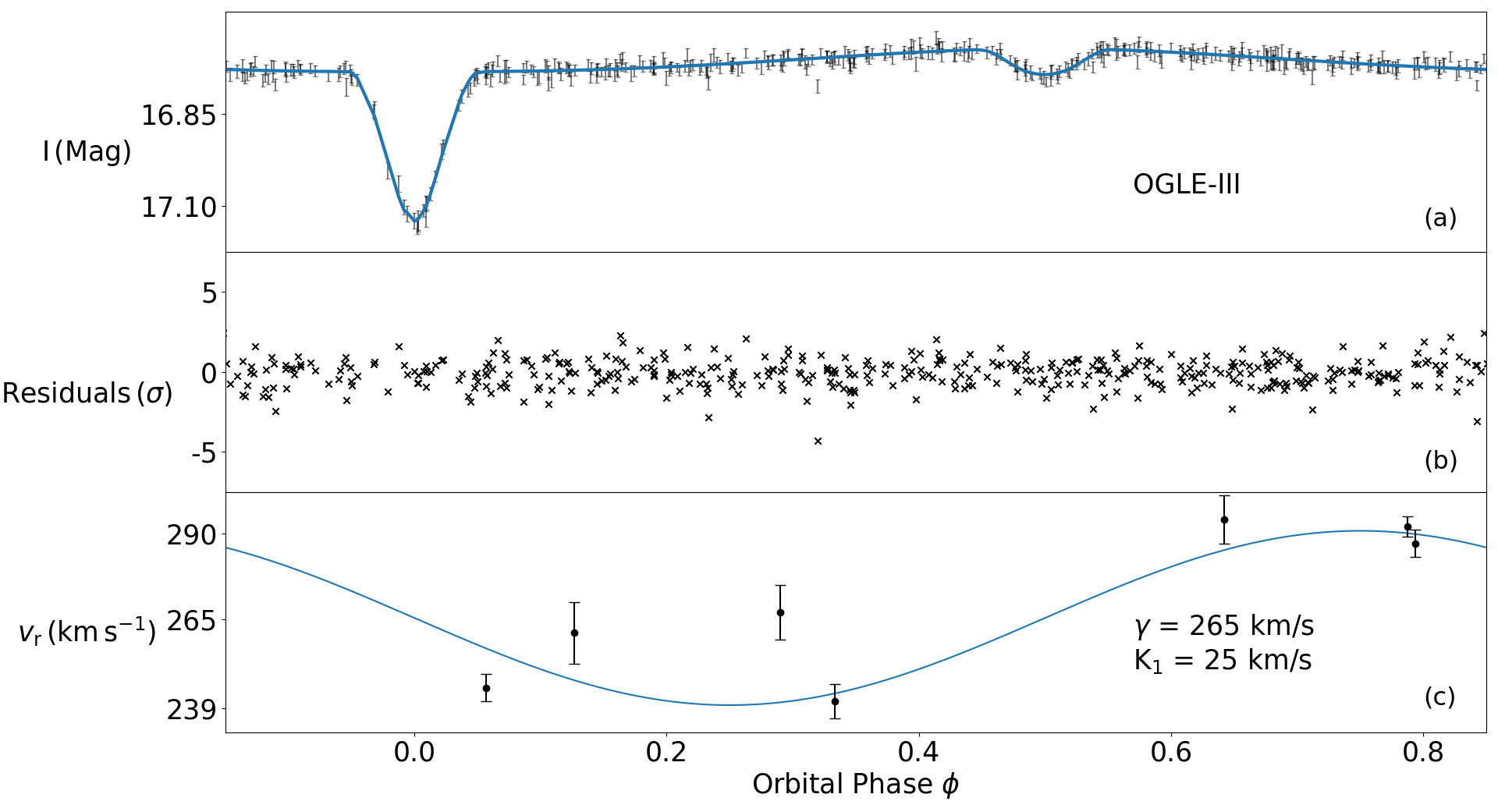}
    \caption{Same as Figure~\ref{fig:rvlc2139} but for EB-21641}
    \label{fig:rvlc21641}
\end{figure*}

We display the best-fit PHOEBE light curve models in Figures~\ref{fig:rvlc2139}, ~\ref{fig:rvlc5481}, and~\ref{fig:rvlc21641} for EB-2139, EB-5481, and EB-21641, respectively. We report the best-fit parameters in Table~\ref{tab:LMFitResults}. We also compute the Roche Lobe Fill Factors (RLFFs) for each component using Equation 2 of~\citet{Eggleton1983}. The fits of our light curves are relatively good with \cs/$\nu$~=~0.77\,-\,1.37. Additionally, there are no systematic features present in the residuals (panel b of Figures~\ref{fig:rvlc2139}, ~\ref{fig:rvlc5481}, and~\ref{fig:rvlc21641}).

As done in \citet{Frank2026}, we add the errors determined from the PDFs and correlated errors in quadrature. For example, PHOEBE precisely measures the temperature ratio $T_{\rm eff,2}$/$T_{\rm eff,1}$, and so we propagate the measured uncertainties in $T_{\rm eff,1}$ when computing the errors in $T_{\rm eff,2}$. Similarly, PHOEBE measures the ratios $R_1/a$ and $R_2/a$, and thus we propagate the uncertainties in $a$, which derive from the uncertainties in $M_1$ and $M_2$ via Kepler's third law.

The measured albedos of our cool secondaries span 45\%\,-\,69\%. \citet{Rucinski1969} determined that cool stars with deep convective envelopes have bolometric albedos of 40\%\,-\,50\%. The consistency between our measured values and the literature supports that the irradiation and reflection effects in our systems were accurately modeled with PHOEBE.

The primary radii $R_\mathrm{1}$ measured via our light curve fits (Table~\ref{tab:LMFitResults}) are systematically smaller than the radii inferred from the SEDs (Table~\ref{tab:SEDfit}). However, the radii from the SEDs appeared to be overinflated for primaries in nascent EBs. Moreover, the radii measured from the SEDs had substantially larger uncertainties. According to MIST evolutionary tracks, the radii of $M_1$ = 6\,-\,9\Msun\ stars on the zero-age MS (ZAMS) span $R_1$ = 2.6\,-\,3.8\,\Rsun. The measured radii $R_1$ = 3.7\,-\,4.4\,\Rsun\ based on the light curve fits are much more consistent with young mid-B primaries near the ZAMS.

\begin{deluxetable*}{lrrrrrrrrrrr}[hpt]
    \tablecaption{PHOEBE light curve fits \label{tab:LMFitResults}}
    \tablehead{
        \colhead{Object} & \colhead{$R_\mathrm{1}$ (\Rsun)} & \colhead{$R_\mathrm{2}$ (\Rsun)} & \colhead{$T_\mathrm{eff, 2}$ (K)} & \colhead{$i$ (\textdegree)} & \colhead{$A_{\mathrm{2}}$(\%)} & \colhead{$N_{\mathrm{I}}$} & \colhead{$N_{\mathrm{C}}$} & \colhead{\cs\,/\,$\nu$} & \colhead{$M_\mathrm{2}$ (\Msun)} & \colhead{$RLFF_{\mathrm{1}}$} & \colhead{$RLFF_{\mathrm{2}}$}
    }
    \startdata
    EB-2139 & 4.15 & 2.00 & 3510 & 86.9 & 69 & 477 & 0 & 1.21 & 0.39 & 0.18 & 0.34 \\
     & $\pm$\,0.13 & $\pm$\,0.07 & $\pm$\,210 & $\pm$\,0.3 & $\pm$\,7  & - & - & - & $\pm$\,0.07 & $\pm$\,0.03 & $\pm$\,0.03 \\
    \hline
    EB-5481 & 4.41 & 2.78 & 4100 & 70.6 & 45 & 847 & 0 & 1.37 & 0.95 & 0.47 & 0.81 \\
     & $\pm$\,0.16 & $\pm$\,0.14 & $\pm$\,320 & $\pm$\,0.5 & $\pm$\,5  & - & - & - & $\pm$\,0.11 & $\pm$\,0.03 & $\pm$\,0.03 \\
    \hline
    EB-21641 & 3.73 & 2.14 & 3980 & 83.2 & 56 & 436 & 0 & 0.77 & 0.60 & 0.36 & 0.62 \\
     & $\pm$\,0.13 & $\pm$\,0.10 & $\pm$\,290 & $\pm$\,0.5 & $\pm$\,6  & - & - & - & $\pm$\,0.07 & $\pm$\,0.03 & $\pm$\,0.03 \\
    \enddata
\end{deluxetable*}

\subsection{Ages and Masses of Pre-MS Secondaries}\label{sec:ages}

\begin{figure}[htb!]
    \centering
    \includegraphics[width=\linewidth]{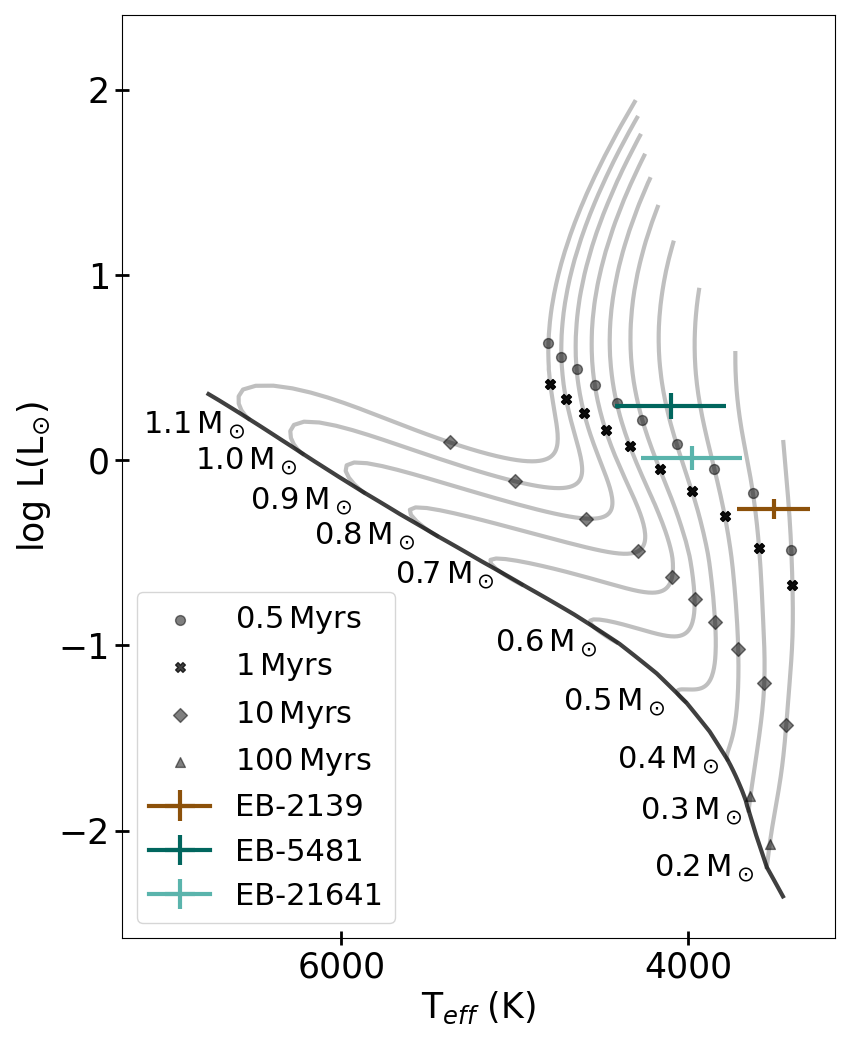}
    \caption{HRD of the pre-MS secondaries in our three nascent EBs. We overlay MIST pre-MS evolutionary tracks and the zero-age MS. The various symbols along each track indicate ages of 0.5, 1, 10, and 100\,Myrs.}
    \label{fig:HRD2}
\end{figure}

\begin{deluxetable*}{lrrrrrrrrrrrrr}[hbt!]
    \setlength{\tabcolsep}{3pt}
    \tablecaption{Finalized System Parameters \label{tab:finalTable}}
    \tablehead{
        \colhead{} & \multicolumn{2}{c}{System} & \colhead{} & \multicolumn{4}{c}{Primary} & \colhead{} & \multicolumn{5}{c}{Secondary} \\
        \colhead{Object} & \colhead{$P$(days)} & \colhead{$i$(\textdegree)} & \colhead{} & \colhead{$M_\mathrm{1}$(\Msun)} & \colhead{$T_\mathrm{eff, 1}$(K)} & \colhead{$R_\mathrm{1}$(\Rsun)} & \colhead{$RLFF_{\mathrm{1}}$} & \colhead{} & \colhead{$M_\mathrm{2}$(\Msun)} & \colhead{$T_\mathrm{eff, 2}$(K)} & \colhead{$R_\mathrm{2}$(\Rsun)} & \colhead{$RLFF_{\mathrm{2}}$}  & \colhead{Age (Myr)} 
    }
    \startdata
    EB-2139 & 8.462510 & 86.9 & & 8.3 & 22000 & 4.15 & 0.18 & & 0.39 & 3510 & 2.00 & 0.34 & 0.9 \\
     & $\pm$\, 0.000015 & $\pm$\,0.3 & & $\pm$\,0.7 & $\pm$\,1200 & $\pm$\,0.13 & $\pm$\,0.03 & & $\pm$\,0.07 & $\pm$\,210 & $\pm$\,0.07 & $\pm$\,0.03 & $\pm$\,0.3 \\
    \hline
    EB-5481 & 2.487130 & 70.6 & & 8.5 & 21800 & 4.41 & 0.47 & & 0.95 & 4100 & 2.78 & 0.81 & 1.0 \\
     & $\pm$\,.000004  & $\pm$\,0.5 & & $\pm$\,0.8 & $\pm$\,1400 & $\pm$\,0.16 & $\pm$\,0.03 & & $\pm$\,0.11 & $\pm$\,320 & $\pm$\,0.14 & $\pm$\,0.03 & $\pm$\,0.4 \\
    \hline
    EB-21641 & 3.092424 & 83.2 & & 6.8 & 18600 & 3.73 & 0.36 & & 0.60 & 3980 & 2.14 & 0.62 & 0.9 \\
     & $\pm$\, 0.000005 & $\pm$\,83.2 & & $\pm$\,0.6 & $\pm$\,800 & $\pm$\,0.13 & $\pm$\,0.03 & & $\pm$\,0.07 & $\pm$\,290 & $\pm$\,0.10 & $\pm$\,0.03 & $\pm$\,0.3 \\
    \enddata
\end{deluxetable*}

We report the main parameters of our nascent EBs in Table~\ref{tab:finalTable}.
We present an HRD with our pre-MS secondaries in Figure~\ref{fig:HRD2}. We overlay MIST pre-MS evolutionary tracks with tick marks indicating age \citep{Choi2016}. The pre-MS masses inferred from the evolutionary tracks are consistent with their dynamical masses (within 1.5$\sigma$). The pre-MS secondaries are located high on the Hayashi track near ages of $\approx$\,1 Myr. We estimate the ages of the pre-MS secondaries from the evolutionary tracks, propagating the uncertainties in $T_{\rm eff,2}$, $L_2$, and $M_2$. We report the ages and uncertainties for our nascent EBs in the final column of Table~\ref{tab:finalTable}. The extreme youth of our nascent EBs suggests they likely hardened via disk migration during the previous embedded protostellar phase \citep{Tokovinin2020, Moe2026}. 

The masses of our pre-MS secondaries span $M_2$ = 0.39\,-\,0.90\,\Msun, yielding extremely small mass ratios $q$ = $M_2$/$M_1$ = 0.05\,-\,0.11. Solar-type primaries exhibit a dearth of close companions below $q$ $<$ 0.1 known as the brown dwarf desert \citep{Grether2006}. In the disk migration model of \citet{Tokovinin2020}, brown dwarf companions to solar-type primaries accrete substantial mass from the circumbinary disk as they migrate inward, ending their migration as close stellar-mass companions. Conversely, migration through the massive protostellar disks of B stars is rather quick and efficient. The model of \citet{Tokovinin2020} produces close companions to B~stars with $q$ $<$ 0.1 (see their Fig.~6). Our nascent EBs with dynamical mass ratios $q$ $<$ 0.1 demonstrate that low-mass companions to massive stars can survive migration to very short periods without accreting much additional mass.

\section{Summary}\label{sec:Summary}

We obtained multi-epoch echelle spectra of three nascent EBs in the LMC with the MIKE spectrograph at the 6.5m Magellan-Clay telescope. By fitting TLUSTY atmospheric models to the normalized spectra, we measured the effective temperatures $T_{\rm eff,1}$, surface gravities log\,$g$, projected rotational velocities $v$\,sin\,$i$, and radial velocities $v_{\rm r}$ of the  B-dwarf primaries. We measured the B-dwarf luminosities $L_1$ and dust extinctions A$_{\rm V}$ by fitting SEDs to publicly available broadband photometry. Based on their locations on an HRD, the primaries are mid-B dwarfs with estimated masses $M_1$ = 6.8\,-\,8.5\,\Msun. In addition to obtaining radial velocities based on the entire spectrum, we fit Gaussian profiles to five deep absorption lines. Our nascent EBs are tidally circularized, and therefore we fit simple sinusoids to the radial velocity curves. We measured the systematic velocities $\gamma$ and velocity amplitudes $K_1$, which yielded the dynamical masses $M_2$\,sin\,$i$ of the pre-MS secondaries. We then fit PHOEBE models to the light curves, constraining five model parameters: radii $R_1$ and $R_2$ of both components, effective temperature $T_{\rm eff,2}$ and albedo $A_2$ of the pre-MS secondary, and inclination $i$ of the orbit.

The pre-MS secondaries are high on the Hayashi track where evolutionary models indicate ages of only 1\,Myr. The extreme youth of nascent EBs suggest they hardened to very short periods via disk migration during the previous protostellar phase. The masses of the pre-MS secondaries are sub-solar, yielding extremely small mass ratios $q$ = 0.05\,-\,0.11. Our nascent EBs demonstrate that low-mass companions to massive stars can migrate through the protostellar disk to very short periods without accreting much additional mass.

\facilities{This work made use of the Magellan Inamori Kyocera Echelle (MIKE) spectrograph attached to the Magellan 2 - Clay Telescope at the Las Campanas Observatory.}

\software{scipy \citep{SciPy}, PHysics Of Eclipsing BinariEs (PHOEBE) Modeling Software \citep{Prsa2005}}

\begin{acknowledgements}

Computations were performed using the University of Wyoming (UW) Advance Research Computing Center MedicineBow HPC, a UW managed computational resource available to UW researchers including faculty, staff, students, and collaborators \citep{UWARCC}.

\end{acknowledgements}

\bibliographystyle{aasjournalv7}  
\bibliography{bib}

\end{document}